# High-performance and reliable probabilistic Ising machine based on simulated quantum annealing

Eleonora Raimondo[1,2], Esteban Garzón[3], Yixin Shao[4], Andrea Grimaldi[5,2], Stefano Chiappini[1], Riccardo Tomasello[6], Noraica Davila-Melendez[7], Jordan A. Katine[7], Mario Carpentieri[6], Massimo Chiappini[1], Marco Lanuzza[3], Pedram Khalili Amiri[4], and Giovanni Finocchio[2*]

[1] *Istituto Nazionale di Geofisica e Vulcanologia, Rome, 00143, Italy*
[2] *Department of Mathematical and Computer Sciences, Physical Sciences and Earth Sciences, University of Messina, Messina, 98166, Italy*
[3] *Department of Computer Engineering, Modeling, Electronics and Systems, University of Calabria, Rende, 87036, Italy*
[4] *Department of Electrical and Computer Engineering, Northwestern University, Evanston, IL, 60208, United States of America*
[5] *Department of Engineering, University of Messina, Messina, 98166, Italy*
[6] *Department of Electrical and Information Engineering, Politecnico di Bari, Bari, 70126, Italy*
[7] *Western Digital, San Jose, CA, 95119, Unites States of America*

[*]Corresponding author: gfinocchio@unime.it

Probabilistic computing with p-bits is emerging as a computational paradigm for machine learning and for facing combinatorial optimization problems (COPs) with the so-called probabilistic Ising machines (PIMs). From a hardware point of view, the key elements that characterize a PIM are the random number generation, the nonlinearity, the network of coupled p-bits, and the energy minimization algorithm. Regarding the latter, in this work we show that PIMs using the simulated quantum annealing (SQA) schedule exhibit better performance as compared to simulated annealing and parallel tempering in solving a number of COPs, such as maximum satisfiability problems, planted Ising problem, and travelling salesman problem. Additionally, we design and simulate the architecture of a fully connected CMOS-based PIM able to run the SQA algorithm having a spin-update time of 8 ns with a power consumption of 0.22 mW. Our results also show that SQA increases the reliability and the scalability of PIMs by compensating for device variability at an algorithmic level enabling the development of their implementation combining CMOS with different technologies such as spintronics. This work shows that the characteristics of the SQA are hardware-agnostic and can be applied in the co-design of any hybrid analog-digital Ising machine implementation. Our results open a promising direction for the implementation of a new generation of reliable and scalable PIMs.

# I. INTRODUCTION

Ising machines (IMs) are emerging as a promising hardware-friendly unconventional computing paradigm for the solution of combinatorial optimization problems (COPs) [1–3], a class of computational problems that have important applications in many different fields, including industry and logistics [4,5]. To solve a COP, IMs search for the ground state of an Ising Hamiltonian, which is defined as

$$H(\boldsymbol{m}) = -\left(\sum_{i<j} J_{ij} m_i m_j + \sum_i h_i m_i\right), \tag{1}$$

where $\boldsymbol{m}$ is the Ising spin vector, with each spin being a binary variable $m_i \in \{-1, +1\}$. $J_{ij}$ is the coupling between the spins $i$ and $j$, and $h_i$ is an external bias acting on the spin $i$. There are several promising paradigms exploiting the hardware implementation of IMs, including quantum annealers [6,7], simulated bifurcation machines [8], coherent [9,10] and oscillator-based IMs [11,12], and probabilistic IMs (PIMs). The latter can be implemented with solutions based on semiconductor technology [13–15], optics [16], spintronics [17–23] and combinations of them [24] and can work at room temperature and used for edge computing. The Ising spin in PIMs is the probabilistic bit (p-bit), its behavior described by an update equation that is composed of a stochastic and a deterministic component. A well-established model for updating a p-bit is

$$m_i(n+1) = \text{sgn}\big(\text{rand}(-1,+1) + \tanh(I_i(n))\big), \tag{2}$$

where $\text{rand}(-1,+1)$ is a random number between –1 and +1, and $I_i(n)$ is the input signal that controls the probability of the $i^{th}$ p-bit being updated in one state or the other via a nonlinear function, i.e., the hyperbolic tangent function (tanh). The binarization process is managed by the sign function. $I_i(n)$ depends on the whole set of $N$ p-bits (p-bit vector state) and how they interact with the $i^{th}$ p-bit. These interactions are described by the **J** matrix, while the local bias is contained in the **h** vector:

$$I_i(n) = \beta \left(\sum_j J_{ij} m_j + h_i\right). \tag{3}$$

Here $\beta$ is the inverse pseudo-temperature, a parameter used for the annealing [1,19,25]. Equation (3) can be seen as a synaptic part, i.e., the synapse, of the update equation as it is similar to the multiply–accumulate operation (MAC) implemented for neurons in neuromorphic computing [26]. The energy landscape, defined by the given **J** and **h**, is sampled by following the Boltzmann distribution when the p-bits are properly updated, e.g., they need to be updated sequentially in a fully connected graph [13].

For a given problem, the identification of **J** and **h** is a key requirement. For example, when a COP can be described by logical circuits, it is possible to map instances of it into an Ising Hamiltonian by employing the strategy of invertible logic circuits, which allow one to use flexible designs based on logical operations that can be operated omnidirectionally [17,19,27]. The truth table of the logic circuit defines the Ising Hamiltonian and its ground state.

Together with the hardware characteristic, the effective search for the ground state of a given Ising Hamiltonian with the PIM is governed by the parameter $\beta$ [28]. Indeed, many energy minimization algorithms, i.e., annealing schemes, tune this parameter with different strategies in order to search for the Ising spin configuration with the lowest energy [1]. Some of the most well-known algorithms for PIMs are: simulated annealing (SA) [14,15], parallel tempering (PT) [19] and simulated quantum annealing (SQA) [20]. Here, we present a comprehensive comparison of these energy minimization approaches on four different COPs – the planted Ising problem (spin glass with planted solution), the maximum satisfiability problem (Max-SAT), weighted Max-SAT (wMax-SAT), and the travelling salesman problem (TSP) (see Appendix A for their brief description). Each comparison also evaluates the performance of the three algorithms as a function of the number of replicas (parallel instances of the p-bit vector state). The results show that SQA performs better than the other annealing schemes. We then simulate a fully connected PIM, implemented using an ASIC-based design with 65 nm CMOS technology, that is able to run the SQA algorithm and we evaluate the energy and area occupancy characteristics of the design.

However, ensuring high-quality random numbers is of primary importance for improving the performance of PIMs. Superparamagnetic magnetic tunnel junctions (MTJs) represent a key pathway in this regard, enabling the potential development of PIMs using a co-integration of CMOS and spintronic technology. The main challenge to overcome here is the intrinsic device-to-device variation of the nonlinear response of the MTJs, meaning that the nonlinearity, i.e. tanh of Eq. (1), can have a slightly different slope for every p-bit. With this in mind, we have performed a systematic study where the range of the slopes to consider has been identified from experimental data of nominally identical MTJs. The SQA significantly outperforms the other annealing schemes, being capable to converge to the optimal solution for a wider range of the "noise" amplitude induced by the device-to-device variations of the MTJs than SA and PT. This co-design approach opens a path for the realization of a more reliable and compact generation of PIMs combining hardware and sophisticated energy minimization algorithms.

## II. ENERGY MINIMIZATION ALGORITHMS

The choice of the annealing schemes and their corresponding hyperparameters can have a great impact on the performance of a PIM. Here, we focus on three annealing schemes, SA, PT, and SQA, and evaluate their performance in addressing four COPs with different topologies, as discussed later. These three energy minimization schemes are briefly introduced below in the text and their main concepts are summarized in FIG. 1. The success probability of solving these problems depends on several factors, including the number of replicas used and the values of the hyperparameters of the algorithm. In this regard, we have performed a massive systematic study varying the number of replicas and using SA, PT, and SQA to evaluate the performance in terms of energy or loss minimization of four different COPs.

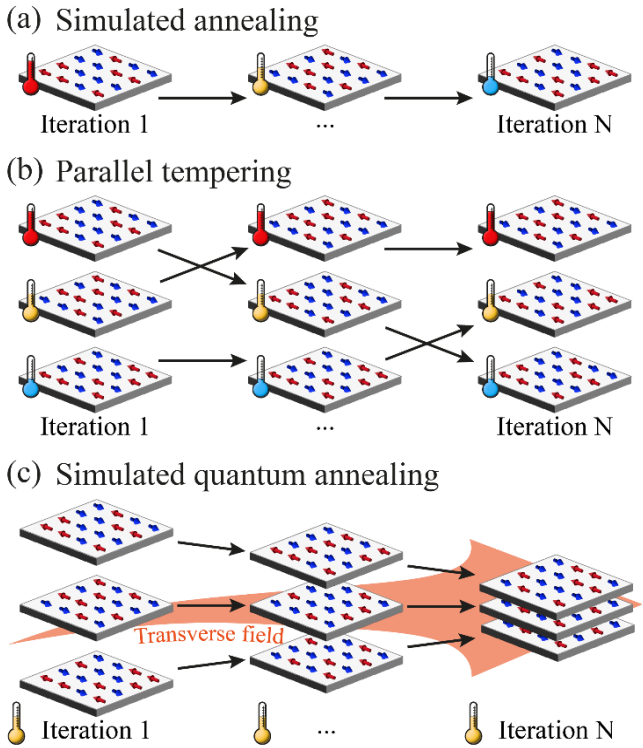

FIG. 1. A schematic representation of the three annealing schemes compared in this work. (a) Simulated annealing: a randomly initialized replica of the system evolves with a linearly decreased temperature. Several replicas can be run in parallel. (b) Parallel tempering: multiple replicas of the system are randomly initialized and evolve at fixed temperature. An operator manages the swap between replicas. (c) Simulated quantum annealing: several random initialized replicas of the system evolve at the same fixed temperature. An increasing transverse field enforces the coupling between replicas, leading to a collective evolution that ideally converges to a single state towards the end of the annealing schedule.

*Simulated annealing.* FIG. 1(a) shows the concept of the SA. At the first iteration, the p-bits configuration is initialized at a random state and the temperature is set at high value ($\beta_H$ close to zero), which promotes the exploration of the spin configuration space. As the number of iterations increases, the temperature linearly decreases up to $\beta_C$ (last iteration). During this time evolution, the spin state evolves toward the lowest-energy states, ideally reaching the ground state. Below a certain temperature, the spin configuration reaches a "frozen" state characterized by an energy barrier larger than the thermal energy. In order to compare SA with PT and SQA, the SA evaluation is based on a number of independent trials equal to the number of replicas, and only the best result among all the trials is selected for the comparison.

*Parallel tempering.* The idea behind PT is shown in FIG. 1(b). Multiple replicas of the spin state are running in parallel, each of them at a different temperature. All the replicas work collectively to search the ground state thanks to a swap operator, see [19] for a more detailed discussion. Higher temperature replicas focus on "diversification", exploring a larger spin configuration space. On the other hand, lower temperature replicas are responsible for "intensification", following the energy gradient of a confined region and finding its local energy minimum [19]. In our study, the inverse temperature of the replicas is linearly increased from $\beta_H$ to $\beta_C$.

*Simulated quantum annealing.* FIG. 1(c) shows the basic concepts of SQA, a quantum-inspired algorithm that simulates quantum tunneling with a transverse field between replicas. The Hamiltonian of the Ising model with the transverse field component can be expressed as [29,30]

$$H(\boldsymbol{m}) = -\sum_{k=1}^{R}\left(\beta\left(\sum_{i<j}J_{ij}m_{i,k}m_{j,k} + \sum_{i}h_i m_{i,k}\right) + J_T\sum_{i}m_{i,k}m_{i,k+1}\right), \qquad (4)$$

where $R$ is the number of replicas, which all have the same temperature $\beta$. The input signal for updating the $i^{th}$ p-bit of the $k^{th}$ replica takes into account the local field computed using equation (3) and the transverse field component, which depends on the $i^{th}$ p-bit state of adjacent replicas:

$$I_i(n) = \beta\left(\sum_{j}J_{ij}m_j + h_i\right) + J_T(m_{i,k-1} + m_{i,k+1}) . \qquad (5)$$

The strength of the transverse field component tunes the interactions among the replicas, and its annealing schedule is given by [20,25]

$$J_T(n) = -J_{T0}\log\left(\tanh\left(\beta\frac{N-n}{N-1}G_x\right)\right), \qquad (6)$$

where $J_{T0}$ is the scaling parameter, $G_x$ is the transverse field, $N$ is the number of iterations, and $n$ is the iteration number of the simulation. At the first iteration, the transverse field is close to zero, allowing the replicas to evolve relatively independently. This field smoothly increases over the course of the simulation, reaching a very large value near the end of the simulation, causing neighboring replicas to be strongly coupled. This coupling, which can be seen as an exchange interaction typical of ferromagnets, drives the stabilization of a single spin vector state or a number of spin vector clusters, each characterized by a different spin configuration. In the latter, we select the cluster with the best configuration.

## III. A COMPARISON OF THE THREE ANNEALING SCHEMES

As already discussed, a comparative analysis of the three annealing schemes, SA, PT, and SQA is based on four different COPs, and it is performed as a function of the number of replicas. This hyperparameter is of crucial importance in the single run of PT and SQA. For SA, it represents the number of runs performed in parallel. The computational analysis involves a heuristic optimization of the parameters for the annealing schemes, summarized in TABLE I, and a maximum number of replicas equal to 1000 (see Supplemental Material Note 1 for additional details [31]). The number of iterations is set to 2000 for the planted Ising instance and 1000 for the Max-SAT, wMax-SAT, and TSP instances.

TABLE I. Summary of the parameters employed for SA, PT and SQA in the simulated instance of COPs.

|  | Simulated annealing | Parallel tempering | Simulated quantum annealing |
|---|---|---|---|
| Planted Ising | $\beta_H = 0; \beta_C = 6$ | $\beta_H = 0.01; \beta_C = 5.0$ | $\beta = 1.0; J_{T0} = 1.0; G_x = 0.5$ |
| Max-Sat | $\beta_H = 0; \beta_C = 1.2$ | $\beta_H = 0.01; \beta_C = 0.5$ | $\beta = 0.5; J_{T0} = 1.0; G_x = 3.0$ |
| wMax-Sat | $\beta_H = 0; \beta_C = 0.2$ | $\beta_H = 0.001; \beta_C = 0.05$ | $\beta = 0.05; J_{T0} = 1.2; G_x = 30.0$ |
| TSP | $\beta_H = 0.03; \beta_C = 0.12$ | $\beta_H = 0.01; \beta_C = 0.065$ | $\beta = 0.045; J_{T0} = 0.58; G_x = 15.0$ |

FIG. 2 shows the results of a comparison among SA, PT and SQA for the four instances of different COPs. The number of replicas ranges from 5 to 1000, and each curve is obtained with the average of the lower envelope of the problem-related metric (*y*-axis) considering 100 runs. The solid line represents the average value at the last iteration, and the shaded area is the standard deviation. Red, green, and blue colors correspond to SA, PT, and SQA, respectively.

Across all the COP instances, a larger number of replicas leads to an improved performance in terms of average number of attempts to reach the optimal solution for each annealing scheme. In particular,

SQA converges to the optimal solution with a lower number of replicas compared to SA and PT. The performance of SA is particularly interesting, as it struggles to find the optimal solution across all the COPs, with the exception of the planted Ising instance, in which SA exhibits high success probability with only few replicas.

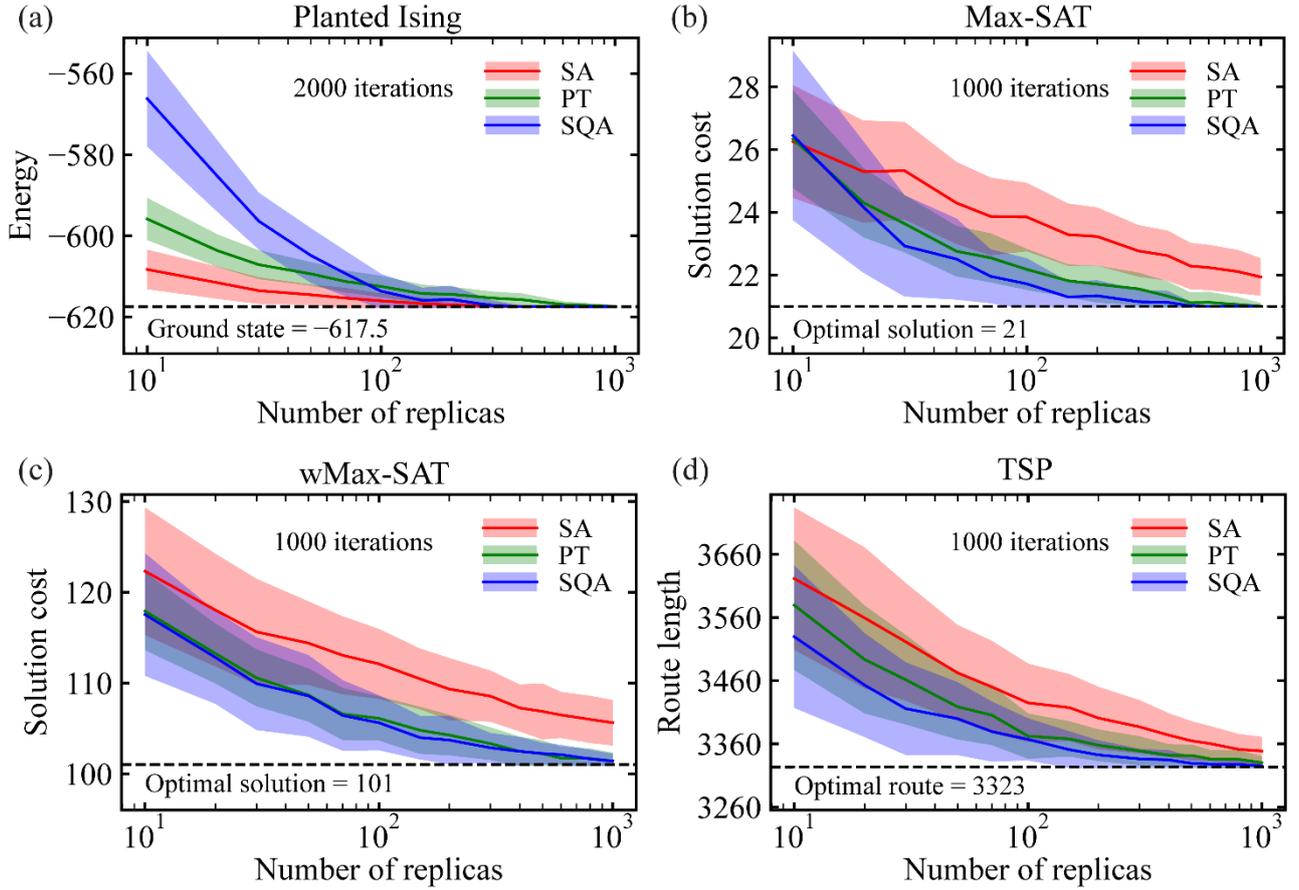

FIG. 2. Results of the comparison among SA (red), PT (green), and SQA (blue) for increasing number of replicas in terms of: (a) energy for a planted Ising instance with a Pegasus topology graph with 1288 p-bits; solution cost for (b) the Max-SAT instance "s3v70c700-1.cnf" with 771 p-bits and (c) the wMax-SAT instance "s3v70c700-1.wcnf" with 771 p-bits; (d) length of the route for the TSP instance "burma14" with 196 p-bits. The solid line is the trend of the average best result at the last iteration (2000 for (a) and 1000 for (b)-(d)) over 100 runs; the area around the solid line is the standard deviation. The dashed black line represents the optimal value for each COPs.

However, as shown in FIG. 3, for a given number of computation time, the SA requires a larger number of iterations to obtain a high success probability in comparison to the SQA – up to 1500 iterations for the SA and up to 500 iterations for the SQA for the examples reported in Fig. 2. Therefore, the slightly additional computational cost of the SQA, related to the computation of the transverse field, is more than justified by the significant reduction in the iterations required for the convergence of the algorithm to the solution. These results are consistent with the theoretical calculations presented in Ref. [32].

It is important to notice that these outcomes are strongly related to the selected parameters, which we tried to optimize here for all the annealing schemes. When searching for the most effective annealing schedule for a fixed number of replicas, only three parameters need to be chosen for SQA, while PT requires setting a value of $\beta$ for each replica. Hence, finding parameters that lead to the optimal solution in SQA appears to be a more straightforward procedure than in PT's case.

We wish to highlight that the identification of the number of replicas for SQA and PT can be chosen considering systematic studies similar to the one reported in Fig. 2 and it is problem dependent.

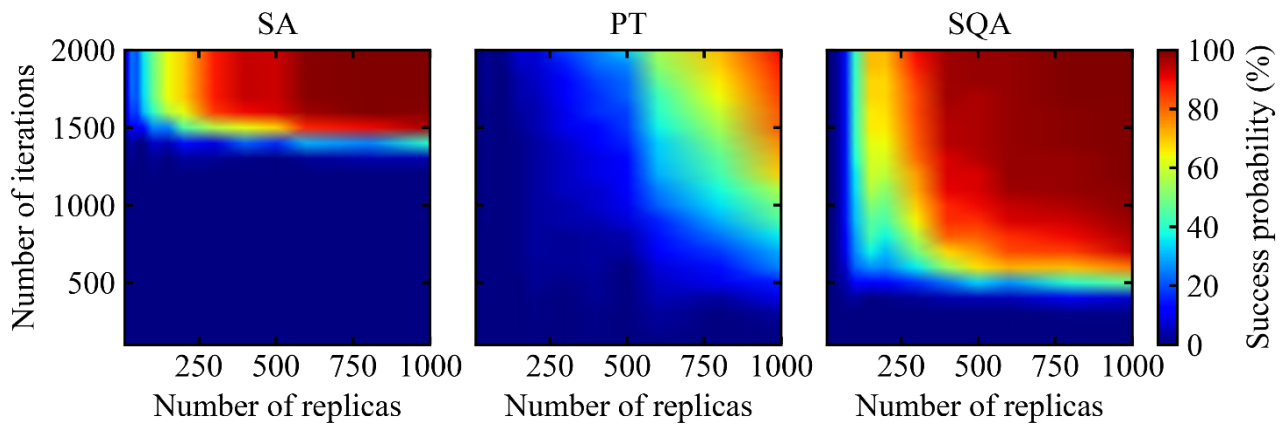

FIG. 3. Comparison of success probability of SA, PT, and SQA at a given number of iterations of a 2000 iterations schedule as a function of the number replicas. The problem considered here is a planted Ising instance with Pegasus graph topology with 1288 p-bits, as the one shown in FIG. 2(a).

## IV. ARCHITECTURE OF A FULLY CONNECTED CMOS-BASED PIM USING SQA

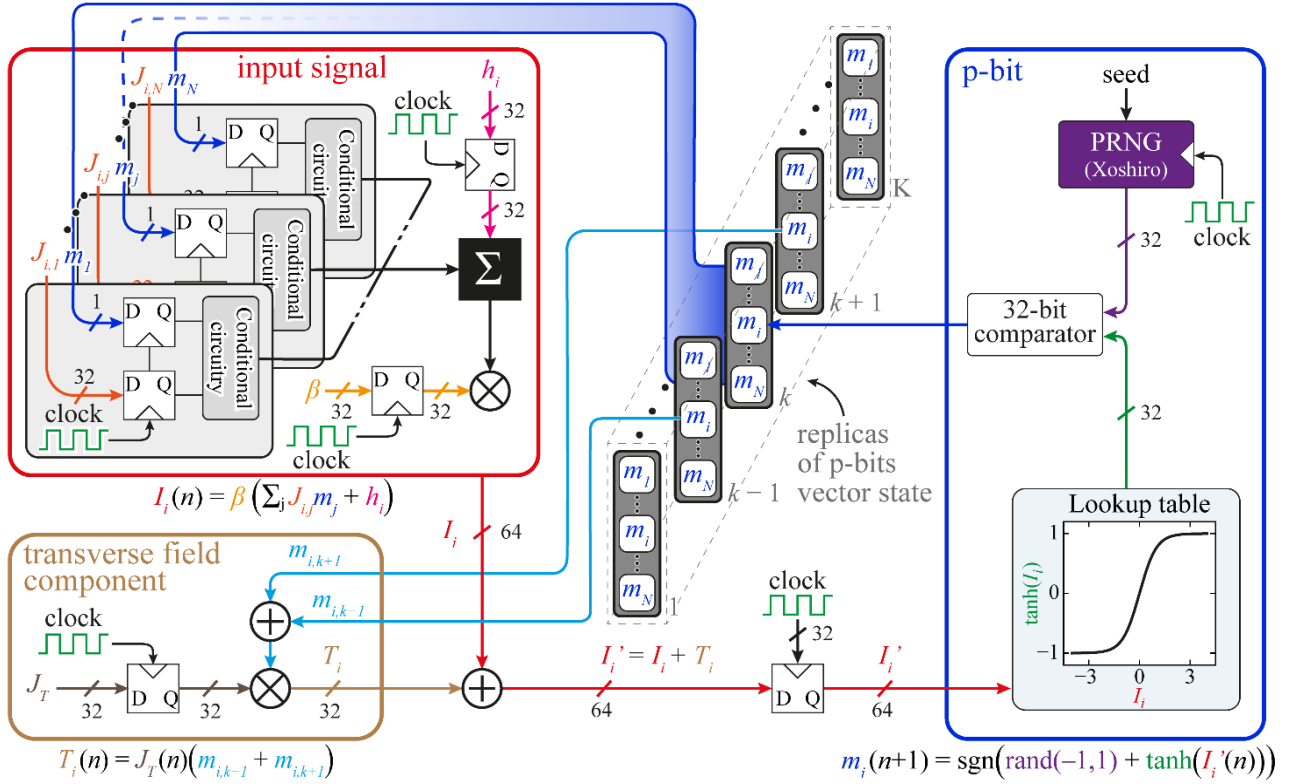

FIG. 4. High-level architecture of the proposed fully connected PIM for SQA. (Top-left) The input signal block consists of a multiply–accumulate operation between the coupling matrix **J** and the vector of p-bits **m**, to which a local bias **h** is added. The resulting value is scaled by the pseudo temperature parameter $\beta$. (Bottom-left) The transverse field component block sums the $i^{th}$ p-bit states of adjacent replicas and scales the result by the transverse field value $J_T$. (Right) The updated state of the $i^{th}$ p-bit is given by a comparison between the tanh of the input signal, computed by a lookup table, and a number generated by the Xoshiro pseudorandom number generator. The updated value is then stored in the memory block of the $i^{th}$ p-bit. Note that, in a fully connected PIM, a single p-bit block is required to serially update the p-bits of the input signal block.

According to the results of the previous section, we designed an SQA-compatible CMOS implementation of PIMs synthesized with Cadence Genus in a commercial 65 nm process design kit (PDK). The top-level architecture, illustrated in FIG. 4, is organized into three main sub-blocks: (1) an input signal block, top left in the figure; (2) a transverse field component block, bottom left in the figure; (3) a p-bit block, right in the figure. The sub-block (1) implements the MAC operation to calculate $I_i(n)$ as defined in Eq. (3). The sub-block (2) takes into account the interaction among different replicas in the SQA algorithm as defined in Eq. (5). The sub-block (3) is used to compute and update the new state of the $i^{th}$ p-bit, $m_i(n + 1)$, based on Eq. (1).

The p-bit is encoded as a binary value, 0 or 1, representing the p-bit values −1 and +1, respectively. A $N\times 1$-bit memory region is used to store the p-bit values of each replica. **J** and **h** are 32-bit fixed point number in 2's complements notation, $\beta$ is a 32-bit fixed point positive value which can be computed offline and then stored in a lookup table (LUT), and each **m** is a $N\times 1$-bit state vector of p-bits. In the input signal block, the product between a vector of **J** and **m** is implemented by a combinatorial conditional circuitry (i.e. without specific arithmetic circuitry). If $m_i = 1$, the $J_{ij}$ remains unchanged, while if $m_i = 0$, the circuitry outputs the 2's complement operation of $J_{ij}$. Afterwards, **h** is summed up with the output of all conditional circuitries, and the result is multiplied by $\beta$. To avoid saturation after this final product operation, the multiplication result, $I_i$, is a 64-bit fixed point number to address the tanh LUT described below.

The transfer field component block is a low-complexity combinational circuitry that, when updating the $i^{th}$ p-bit of the $k^{th}$ replica, receives the state of the $i^{th}$ p-bit from both the previous replica, $m_{i,k-1}$, and the next replica, $m_{i,k+1}$, as well as a 32-bit fixed point number from a LUT. If the $m_{i,k+1} \neq m_{i,k-1}$, the transverse field component, $T_i$, is zero. Otherwise, if $m_{i,k+1} = m_{i,k-1}$ (both +1 or both −1), $J_i$ outputs twice the transverse field value, $2J_T$ or $-2J_T$. Accordingly, the combinatorial circuit was implemented.

The p-bit block is composed of a pseudorandom number generator (PRNG) and a tanh LUT, as shown in FIG. 4**Errore. L'origine riferimento non è stata trovata.**. The tanh LUT takes the 64-bit input $I_i$, and outputs a 32-bit value, approximating the mathematical function with a resolution of 512 discrete values between −1 and +1. The LUT stores 256 discrete 32-bit values, which enables the representation of positive tanh outputs. To handle the bipolar nature of the tanh function, the sign of the input $I_i$ is analyzed separately to determine whether the retrieved LUT value corresponds to a positive or negative result. This sign-magnitude scheme avoids the need to explicitly store negative tanh values, thereby reducing the hardware complexity/resources. The PRNG generates a 32-bit random number between −1 and +1 on each positive edge of the clock using an XOR-shift-rotate (Xoshiro) implementation [33]. Xoshiro-based PRNG is known for its superior statistical properties as compared to a linear-feedback shift register (LFSR) based PRNG [34], at the cost of increased hardware complexity and more additional consumption (about 6%). The outputs from the PRNG and the tanh LUT are compared to return the new state of the p-bit value, which is stored in the p-bit vector state and sent to the input signal block. The p-bits are updated sequentially and synchronously to a digital clock signal, as the PIM was designed to host fully connected graphs, which do not allow parallel updates of the spins. The update of each p-bit requires two clock cycles. The whole circuit is synthesized with a supply voltage of 1.2 V, operating temperature of 25 °C, and a system clock of 4 ns (250MHz).

FIG. 5 shows power consumption and silicon area occupation as a function of the number of p-bits. It can be observed that the power consumption per p-bit per update saturates at about 0.22 mW as the number of p-bits increases, mainly due to the significant power consumption of the MAC operation within the input signal block. The total energy consumption and the runtime for the COPs studied in this work are reported in TABLE II. As a projection, we also estimate the power consumption of the

PIM across various technology nodes using the scaling equations introduced in Ref. [35]. This PIM design exhibits a power consumption per p-bit per update of approximately 59.8 µW, 14.2 µW, and 7.8 µW, for the 32 nm, 16 nm, and 7 nm nodes, respectively. As expected, the larger area footprint (refer to the sequential update curve) of the PIM is mainly attributed to the input signal block (MAC operation) circuitry. For instance, the total area occupied by the PIM with 10, 100, and 500 p-bits is about 0.041 mm$^2$, 0.113 mm$^2$, and 0.429 mm$^2$, respectively, where the area occupancy of the input signal block is 52%, 83%, 97% of the total area, respectively. In Supplemental Material Note 2 [31], we show a performance comparison with digital and synchronous CMOS-based Ising machines [8,14,36–46].

We wish to highlight one more time that this architecture is an IM implementation allowing one to handle both fully connected problems and sparse graph of any topology. However, often the COPs mapped into the Ising Hamiltonian are not fully connected, or the hardware architecture allow for using graphs with a reduced connectivity such as Pegasus, Zephyr, or others [14,47,48]. In these cases, it is possible to design architectures which can perform parallel updates of clusters of p-bits (see Ref. [13]).

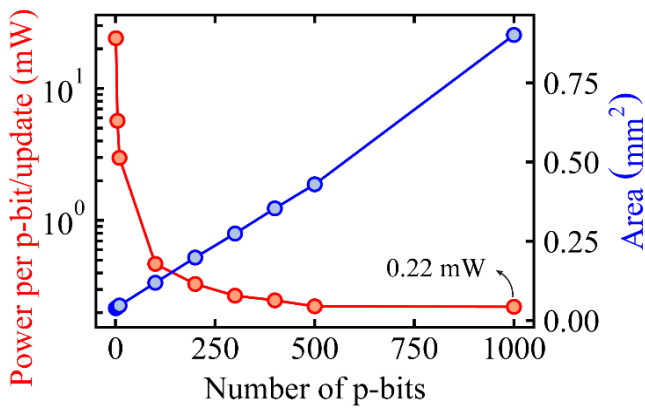

FIG. 5. Power consumption per p-bit per update and area occupancy of the PIM as a function of the number of p-bits which can be used to implement SQA. Saturation is reached around 500 p-bits, exhibiting a power consumption of about 0.22 mW.

TABLE II: Total energy consumption and total runtime of SQA-based PIMs for the COPs instances considering 1000 replicas.

|  | Number of p-bits | Number of iterations | Total energy consumption (mJ) | Total runtime (s) |
|---|---|---|---|---|
| Planted Ising | 1288 | 2000 | 2.28 | 20.6 |
| Max-Sat and wMax-Sat | 771 | 1000 | 0.69 | 6.2 |
| TSP | 196 | 1000 | 0.26 | 1.6 |

## V. MTJ-BASED P-BIT AND ARCHITECTURE OF A HYBRID CMOS-SPINTRONIC BASED PIM

A direction to reduce power consumption and area occupancy is the realization of hybrid PIMs combining CMOS with other technologies such as photonics [49], ferroelectrics [50], spintronics [51]. Focusing on spintronic technology, one effective approach is the implementation of p-bits with MTJs which has also the advantage to work as true RNG (TRNG) [21,52,53]. The design of hybrid PIMs studied here involves the use of MTJs for the implementation of the p-bit block, including both the TRNG and the calculation of the nonlinearity, i.e., tanh, in the same device [13,17,19,20,26,27,34,54,55]. Several different strategies have been proposed to realize MTJ-based TRNGs. Among these, superparamagnetic MTJs offer a promising approach [26,27], encouraged by recent results on nanosecond scale generation of random bits [19,55–58]. The power dissipation of p-bits implemented with superparamagnetic MTJs is nearly independent of the bit generation rate, since the sensing current applied to the MTJ and the corresponding currents and voltages in the associated CMOS circuitry do not change. As a result, since the power dissipation is nearly the same, the energy per generated bit is smaller when working with faster superparamagnetic MTJ devices. Hence, the overall energy dissipated for solving a particular problem reduces as the superparamagnetic MTJs become faster. The tanh response can be also preserved in superparamagnetic MTJs working at nanosecond time scale [57–60]. In particular, the main idea here is to use a single MTJ and three transistors for the realization of a p-bit [13,55].

To identify the main challenges of such an implementation, we consider the device analog characteristics of the experimental superparamagnetic MTJs developed in [61]. This perpendicular MTJ has a circular layout with a nominal diameter of 50 nm and the stack, sketched in FIG. 6(a), consists of a bottom electrode Co/Pt-based synthetic antiferromagnetic multilayer / W coupling layer / CoFeB reference layer / MgO (1.5 nm) / CoFeB free layer (1.4 nm) / top electrode, where the values in parentheses present the thickness of the layer (see Appendix B for more details). The tunnel magnetoresistance ratio is ≈ 130%. At room temperature, the MTJ resistance, which depends on the $z$-component of the magnetization, switches between the parallel and antiparallel states stochastically [61]. The p-bit state can be set by evaluating the resistance of the MTJ at timescales larger than the one governing the thermal dynamics of the magnetization across an energy barrier $E_b$.

The superparamagnetic behavior of the MTJs is tunable via the spin-transfer torque (STT) generated by the applied bias voltage. The resistance subjected to different voltages and a fixed external magnetic field of −35 mT is measured for ≈ 2 min, in intervals of 100 ms, resulting in ≈ 1200 data points for each voltage. Experimental results demonstrate that a bias voltage of less than 1 V achieves a remarkable tunability, enabling transitions from > 95% parallel to > 95% antiparallel states. FIG. 6(b) shows the experimental data (markers) that represent the probability of MTJs being in parallel states ($m_z = +1$) for three different MTJs. The data are well-fitted with a hyperbolic tangent function (solid lines) characterized by a slope coefficient $\alpha$ approximately equal to 2, along with an offset. The offset can be removed through a proper calibration of the device, fixing the bias voltage to have a probability near 50% of the p-bit states. On the other hand, the slope coefficient can be integrated into the parameter $\beta$ by scaling its values. According to these considerations, the slight difference observed among different MTJ curves is basically reflected in the variations of the slope of the tanh function. Similar responses can be observed for other MTJs [62]. The first point to address in view of a future design of these hybrid CMOS-spintronic PIMs is the evaluation of the impact of the device-to-device variation to the performance when compared to ideal solutions, as discussed in the next paragraph.

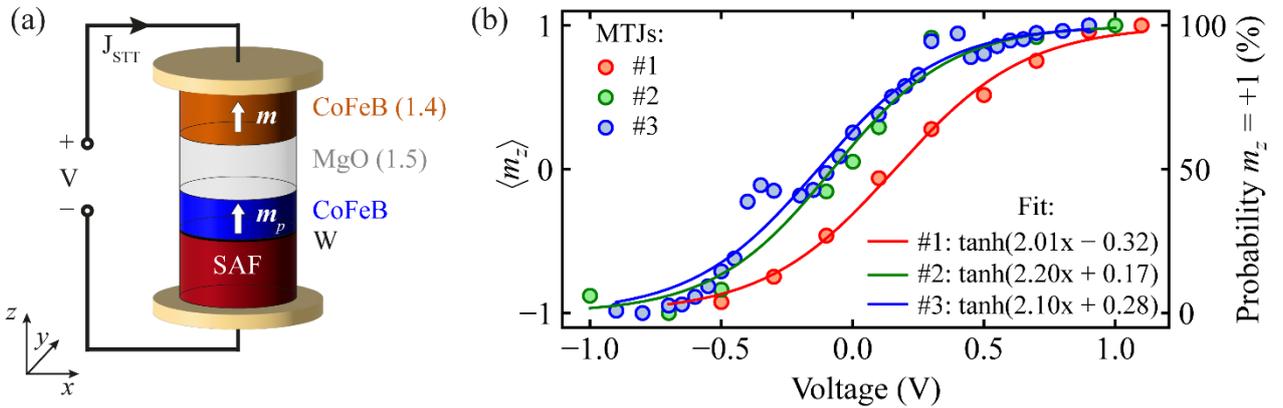

FIG. 6. (a) A sketch of the two-terminal perpendicular MTJ designed for the hardware p-bit implementation. The thickness of the layers in nm is given in parentheses. (b) Probabilities of parallel states ($m_z = +1$) generated by three MTJs (red, green, and blue dots) under different bias voltage. The experimental data are well-fitted with hyperbolic tangent functions (red, green, and solid lines, respectively).

## VI. EVALUATION OF RELIABILITY OF NON-IDEAL PIMS WITH DIFFERENT ANNEALING SCHEMES

As already discussed in section V, the slope variation of non-ideal p-bits can be estimated directly from experimental data. To model these device-to-device variations, the update equation of the $i^{th}$ p-bit becomes $m_i(n+1) = \text{sgn}(\text{rand}(-1,+1) + \tanh(\alpha I_i(n)))$, where $\alpha$ is a p-bit-dependent parameter that is sampled from a Gaussian distribution with unitary mean and a standard deviation $\sigma$.

For realistic assumptions, the α values are confined within the range $[1-\sigma, 1+\sigma]$ as shown in the inset in FIG. 7(a). The ideal PIM has σ equal to zero (α always equal to one). The largest variability considered for the simulations is σ = 0.8, meaning the slope of the p-bits varies from 0.2 to 1.8. Reliability assessments of the PIM in terms of success probability considering 1000 runs for all the three annealing schemes are reported in FIG. 7. The parameters used for the annealing schemes are summarized in TABLE I.

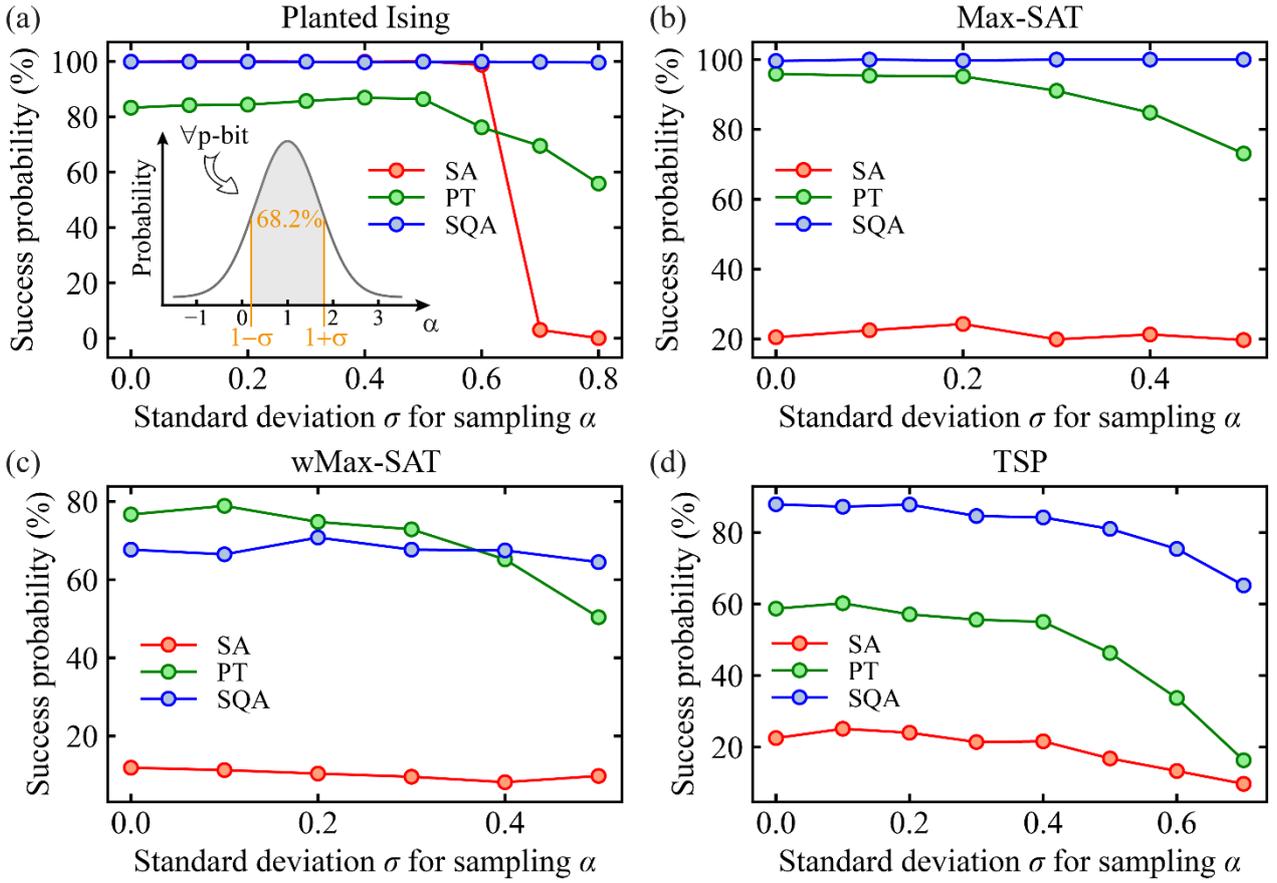

FIG. 7. Success probability over 1000 runs per point as a function of increasing variability in: (a) planted Ising instance with Pegasus topology graph; (b) Max-SAT instance "s3v70c700-1.cnf"; (c) wMax-SAT instance "s3v70c700-1.wcnf"; (d) TSP instance "burma14". Each problem is tested with three energy minimization algorithms, simulated annealing (SA, red), parallel tempering (PT, green), and simulated quantum annealing (SQA, blue). Each algorithm uses 1000 replicas.

In Fig. 7(a)-(d), SA (red), PT (green), and SQA (blue) are tested on one instance of planted Ising, Max-SAT, wMax-SAT, and TSP, respectively (the same annealing schemes considered for the results of Fig. 2). Each algorithm uses 1000 replicas per attempt, each value of the standard deviation is attempted 1000 times. These results reveal that SQA outperforms both PT and SA, exhibiting remarkable insensitivity to the device-to-device variations, positioning it as a robust solution for practical applications.

The SA, although robust to variations, exhibits a low success probability (less than 25%) for instances of Max-SAT and TSP, limiting its usability for practical applications. In the planted Ising instance, SA starts with a high success probability that dramatically decreases under conditions of high device variability.

PT, on the other hand, shows consistently good performance in ideal conditions but greater susceptibility to the slope variations. However, it is noteworthy that, for small values of σ, the impact is only marginal, leading to the conclusion the PT indeed exhibits some degree of robustness.

Finally, SQA shows not only impressive ideal performance (higher than PT in three of the four problems considered), but also a remarkable robustness to even extreme degrees of variability in the slope of the p-bits. For planted Ising, Max-SAT, and wMax-SAT, SQA shows little to no decrease in performance even with σ = 0.8. For TSP, whose encoding is characterized by hard constraints that makes it very susceptible to thermal fluctuations, only causes a decrease in performance of about ≈ 25%. A plausible explanation for this robustness lies in the nontrivial interactions among replicas. Rather than simply forcing the collapse into a single state, as ideally envisioned by quantum inspiration of SQA, the transverse field directly influences the dynamics of the replicas, collectively guiding them toward deep energy minima.

## VII. SUMMARY AND CONCLUSIONS

In this work, we have performed a comprehensive study on how different energy minimization algorithms perform in PIMs when used to solve different COPs. The main result is that the SQA exhibits better performance than SA and PT in most of the problems studied. With this in mind, we have designed and simulated an ASIC-based design with 65 nm CMOS technology of a fully connected PIM that can run the SQA algorithm. The energy and area occupancy characteristics of this design are of the same order of magnitude as the ones reported in literature (see Table II of Ref [40]). Energy improvements can be achieved with solutions working with in-memory computing and near-memory computing approaches. On the other hand, hybrid architectures combining CMOS with other technologies such as spintronics and/or mixed analog-digital computing schemes can be promising directions to reduce the power consumption of PIMs. Concerning these hybrid architectures, we have found that SQA can be used as an energy minimization algorithm that can make PIMs robust against intrinsic device-to-device variations.

**ACKNOWLEDGMENTS**

This work has been supported by project number 101070287 — SWAN-on-chip — HORIZON-CL4-2021-DIGITAL-EMERGING-01, the projects PRIN 2020LWPKH7 "The Italian factory of micromagnetic modeling and spintronics", PRIN_20225YF2S4 "Magneto-Mechanical Accelerometers, Gyroscopes and Computing based on nanoscale magnetic tunnel junctions (MMAGYC)", PRIN20222N9A73 "SKYrmion-based magnetic tunnel junction to design a temperature SENSor—SkySens", PON Capitale Umano (CIR_00030), funded by the Italian Ministry of University and Research (MUR), and by the PETASPIN association (www.petaspin.com). R.T. and M.C. acknowledge support from the Project PE0000021, "Network 4 Energy Sustainable Transition – NEST", funded by the European Union – NextGenerationEU, under the National Recovery and Resilience Plan (NRRP), Mission 4 Component 2 Investment 1.3 - Call for tender No. 1561 of 11.10.2022 of Ministero dell'Università e della Ricerca (MUR) (CUP C93C22005230007). The work at Northwestern University was supported by the National Science Foundation (NSF), under award numbers 2322572 and 2425538.

**APPENDIX A: COMBINATORIAL OPTIMIZATION PROBLEMS**

In order to perform a comprehensive comparison among different annealing schemes, we focus on four instances of NP-hard problems with different topologies and sizes as described in detail below.

*Planted Ising*. The planted Ising problem refers to a spin glass problem that is carefully constructed upon a known solution. These problems are generated by combining multiple small Hamiltonians, each with known ground-state configurations, on the same graph. These configurations are summed together, creating a nontrivial problem with a known, "planted", solution [15,63]. This process allows for the generation of numerous instances of the problem with different size and complexity, making them particularly suitable as benchmarks. Several lattices structures, including Zephyr, Chimera and Pegasus [47], are commonly used in literature, each of these characterized by different graph topologies. In this work, we focus on a Pegasus topology graph instance, with 1288 p-bits.

*Maximum satisfiability problem* (Max-SAT). The Max-SAT is defined by propositional formulas in conjunctive normal form (CNF). In CNF, a propositional formula is a conjunction of clauses, where each clause is a disjunction of literals. For a Boolean variable $x$, a literal is defined as $x$ or its negation -$x$. The goal of Max-SAT is to find an assignment of the variables that maximizes the number of satisfied clauses [64]. Here, we measure the "solution cost" which represents the number of unsatisfied clauses. Each clause is assigned a weight of one, but a more general version of Max-SAT, the weighted Max-SAT (wMax-SAT), assigns a non-negative weight to the clauses, and the goal is to find an assignment that maximizes the combined weight of the satisfied clauses. A Max-SAT instance can be represented as a logic circuit, where literals in a clause are connected by OR gates, and all clauses are connected by AND gates. The regular gates are replaced with their equivalent invertible gates, and the instance is then mapped into the Ising model elements [19]. In this work, we select a Max-SAT and a wMax-SAT instance of average difficulty from the international Max-SAT competition 2016, namely "s3v70c700-1.cnf" and s3v70c700-1.wcnf", respectively. These instances are characterized by 70 variables and 700 clauses (with 3 literals for each clause) and are encoded with 771 p-bits.

*Travelling salesman problem* (TSP). The TSP aims at finding the shortest route to visit $N$ cities once and only once and return to the original city. The TSP can be mapped onto the Ising model by encoding information about the cities and their route order as a spin variable. The total number of spins needed corresponds to the square of the number of cities [3]. Here, we study a well-known benchmark TSP instance with 14 nodes known as "burma14", encoded with 196 p-bits.

**APPENDIX B: MTJ DEVICE FABRICATION AND MEASUREMENTS**

The MTJ stacks were deposited using sputtering on thermally oxidized silicon wafers in an ultrahigh-vacuum physical vapor deposition system. The films were subsequently fabricated into circular pillars with a diameter of 50 nm connected to metallic measurement pads, using electron beam lithography and established fabrication techniques reported in previous works [21,61,65]. Electrical measurements of the patterned devices were performed on a probe station with a projected field electromagnet, using high-frequency probes with a ground-signal-ground (GSG) configuration. Stochastic bit-streams were generated experimentally by measuring the resistance of the MTJs in time domain under different voltage bias conditions, and were subsequently converted into probability curves as indicated in the main text.

# REFERENCES

[1] N. Mohseni, P. L. McMahon, and T. Byrnes, *Ising Machines as Hardware Solvers of Combinatorial Optimization Problems*, Nat. Rev. Phys. **4**, 363 (2022).

[2] Y. Fu and P. W. Anderson, *Application of Statistical Mechanics to NP-Complete Problems in Combinatorial Optimisation*, J. Phys. A. Math. Gen. **19**, 1605 (1986).

[3] A. Lucas, *Ising Formulations of Many NP Problems*, Front. Phys. **2**, 1 (2014).

[4] M. R. Bartolacci, L. J. LeBlanc, Y. Kayikci, and T. A. Grossman, *Optimization Modeling for Logistics: Options and Implementations*, J. Bus. Logist. **33**, 118 (2012).

[5] H. M. V. Samani and A. Mottaghi, *Optimization of Water Distribution Networks Using Integer Linear Programming*, J. Hydraul. Eng. **132**, 501 (2006).

[6] T. Albash and D. A. Lidar, *Demonstration of a Scaling Advantage for a Quantum Annealer over Simulated Annealing*, Phys. Rev. X **8**, 031016 (2018).

[7] A. D. King et al., *Quantum Critical Dynamics in a 5,000-Qubit Programmable Spin Glass*, Nat. 2023 6177959 **617**, 61 (2023).

[8] K. Tatsumura, M. Yamasaki, and H. Goto, *Scaling out Ising Machines Using a Multi-Chip Architecture for Simulated Bifurcation*, Nat. Electron. **4**, 208 (2021).

[9] T. Honjo et al., *100,000-Spin Coherent Ising Machine*, Sci. Adv. **7**, 952 (2021).

[10] A. Litvinenko, R. Khymyn, V. H. González, R. Ovcharov, A. A. Awad, V. Tyberkevych, A. Slavin, and J. Åkerman, *A Spinwave Ising Machine*, Commun. Phys. **6**, 227 (2023).

[11] S. Dutta, A. Khanna, A. S. Assoa, H. Paik, D. G. Schlom, Z. Toroczkai, A. Raychowdhury, and S. Datta, *An Ising Hamiltonian Solver Based on Coupled Stochastic Phase-Transition Nano-Oscillators*, Nat. Electron. **4**, 502 (2021).

[12] J. Vaidya, R. S. Surya Kanthi, and N. Shukla, *Creating Electronic Oscillator-Based Ising Machines without External Injection Locking*, Sci. Rep. **12**, 981 (2022).

[13] S. Chowdhury et al., *A Full-Stack View of Probabilistic Computing With p-Bits: Devices, Architectures, and Algorithms*, IEEE J. Explor. Solid-State Comput. Devices Circuits **9**, 1 (2023).

[14] N. A. Aadit, A. Grimaldi, M. Carpentieri, L. Theogarajan, J. M. Martinis, G. Finocchio, and K. Y. Camsari, *Massively Parallel Probabilistic Computing with Sparse Ising Machines*, Nat. Electron. **5**, 460 (2022).

[15] N. A. Aadit, A. Grimaldi, G. Finocchio, and K. Y. Camsari, *Physics-Inspired Ising Computing with Ring Oscillator Activated p-Bits*, in *2022 IEEE 22nd International Conference on Nanotechnology (NANO)*, Vols. 2022-July (IEEE, 2022), pp. 393–396.

[16] C. Roques-Carmes, Y. Salamin, J. Sloan, S. Choi, G. Velez, E. Koskas, N. Rivera, S. E. Kooi, J. D. Joannopoulos, and M. Soljačić, *Biasing the Quantum Vacuum to Control Macroscopic Probability Distributions*, Science. **381**, 205 (2023).

[17] K. Y. Camsari, R. Faria, B. M. Sutton, and S. Datta, *Stochastic P-Bits for Invertible Logic*, Phys. Rev. X **7**, 031014 (2017).

[18] G. Finocchio, M. Di Ventra, K. Y. Camsari, K. Everschor-Sitte, P. Khalili Amiri, and Z. Zeng, *The Promise of Spintronics for Unconventional Computing*, J. Magn. Magn. Mater. **521**,

167506 (2021).

[19] A. Grimaldi, L. Sánchez-Tejerina, N. Anjum Aadit, S. Chiappini, M. Carpentieri, K. Camsari, and G. Finocchio, *Spintronics-Compatible Approach to Solving Maximum-Satisfiability Problems with Probabilistic Computing, Invertible Logic, and Parallel Tempering*, Phys. Rev. Appl. **17**, 024052 (2022).

[20] A. Grimaldi et al., *Experimental Evaluation of Simulated Quantum Annealing with MTJ-Augmented p-Bits*, in *2022 International Electron Devices Meeting (IEDM)*, Vols. 2022-Decem (IEEE, 2022), pp. 22.4.1-22.4.4.

[21] Y. Shao, C. Duffee, E. Raimondo, N. Davila, V. Lopez-Dominguez, J. A. Katine, G. Finocchio, and P. Khalili Amiri, *Probabilistic Computing with Voltage-Controlled Dynamics in Magnetic Tunnel Junctions*, Nanotechnology **34**, 495203 (2023).

[22] A. Grimaldi, L. Mazza, E. Raimondo, P. Tullo, D. Rodrigues, K. Y. Camsari, V. Crupi, M. Carpentieri, V. Puliafito, and G. Finocchio, *Evaluating Spintronics-Compatible Implementations of Ising Machines*, Phys. Rev. Appl. **20**, 024005 (2023).

[23] J. Si et al., *Energy-Efficient Superparamagnetic Ising Machine and Its Application to Traveling Salesman Problems*, Nat. Commun. **15**, 3457 (2024).

[24] P. L. McMahon et al., *A Fully Programmable 100-Spin Coherent Ising Machine with All-to-All Connections*, Science. **354**, 614 (2016).

[25] A. Grimaldi, E. Raimondo, A. Giordano, K. Y. Çamsarı, and G. Finocchio, *A Comparison of Energy Minimization Algorithms for Solving Max-Sat Problem with Probabilistic Ising Machines*, in *2023 IEEE 23rd International Conference on Nanotechnology (NANO)*, Vols. 2023-July (IEEE, 2023), pp. 698–702.

[26] A. Z. Pervaiz, S. Datta, and K. Y. Camsari, *Probabilistic Computing with Binary Stochastic Neurons*, in *2019 IEEE BiCMOS and Compound Semiconductor Integrated Circuits and Technology Symposium (BCICTS)* (IEEE, 2019), pp. 1–6.

[27] W. A. Borders, A. Z. Pervaiz, S. Fukami, K. Y. Camsari, H. Ohno, and S. Datta, *Integer Factorization Using Stochastic Magnetic Tunnel Junctions*, Nature **573**, 390 (2019).

[28] B. A. Cipra, *An Introduction to the Ising Model*, Am. Math. Mon. **94**, 937 (1987).

[29] M. Suzuki, *Relationship between D-Dimensional Quantal Spin Systems and (D+1)-Dimensional Ising Systems: Equivalence, Critical Exponents and Systematic Approximants of the Partition Function and Spin Correlations*, Prog. Theor. Phys. **56**, 1454 (1976).

[30] T. Okuyama, M. Hayashi, and M. Yamaoka, *An Ising Computer Based on Simulated Quantum Annealing by Path Integral Monte Carlo Method*, in *2017 IEEE International Conference on Rebooting Computing (ICRC)*, Vols. 2017-Janua (IEEE, 2017), pp. 1–6.

[31] See Supplementary Material for a performance comparison of CMOS-based PIM with other Ising machines, as well as details on parameter optimization.

[32] E. Crosson and A. W. Harrow, *Simulated Quantum Annealing Can Be Exponentially Faster Than Classical Simulated Annealing*, in *2016 IEEE 57th Annual Symposium on Foundations of Computer Science (FOCS)*, Vols. 2016-Decem (IEEE, 2016), pp. 714–723.

[33] D. Blackman and S. Vigna, *Scrambled Linear Pseudorandom Number Generators*, ACM Trans. Math. Softw. **47**, 1 (2021).

[34] N. S. Singh et al., *CMOS plus Stochastic Nanomagnets Enabling Heterogeneous Computers*

*for Probabilistic Inference and Learning*, Nat. Commun. **15**, 2685 (2024).

[35] A. Stillmaker and B. Baas, *Scaling Equations for the Accurate Prediction of CMOS Device Performance from 180 Nm to 7 Nm*, Integration **58**, 74 (2017).

[36] R. Iimura, S. Kitamura, and T. Kawahara, *Annealing Processing Architecture of 28-Nm CMOS Chip for Ising Model With 512 Fully Connected Spins*, IEEE Trans. Circuits Syst. I Regul. Pap. **68**, 5061 (2021).

[37] K. Kawamura et al., *Amorphica: 4-Replica 512 Fully Connected Spin 336MHz Metamorphic Annealer with Programmable Optimization Strategy and Compressed-Spin-Transfer Multi-Chip Extension*, in *2023 IEEE International Solid- State Circuits Conference (ISSCC)* (IEEE, 2023), pp. 42–44.

[38] H. M. Waidyasooriya and M. Hariyama, *Highly-Parallel FPGA Accelerator for Simulated Quantum Annealing*, IEEE Trans. Emerg. Top. Comput. **9**, 2019 (2021).

[39] S. Xie, S. R. S. Raman, C. Ni, M. Wang, M. Yang, and J. P. Kulkarni, *Ising-CIM: A Reconfigurable and Scalable Compute Within Memory Analog Ising Accelerator for Solving Combinatorial Optimization Problems*, IEEE J. Solid-State Circuits **57**, 3453 (2022).

[40] Y. Su, H. Kim, and B. Kim, *CIM-Spin: A Scalable CMOS Annealing Processor With Digital In-Memory Spin Operators and Register Spins for Combinatorial Optimization Problems*, IEEE J. Solid-State Circuits **57**, 2263 (2022).

[41] P. Meinerzhagen, A. Teman, R. Giterman, N. Edri, A. Burg, and A. Fish, *Gain-Cell Embedded DRAMs for Low-Power VLSI Systems-on-Chip* (Springer International Publishing, Cham, 2018).

[42] E. Garzon, Y. Greenblatt, O. Harel, M. Lanuzza, and A. Teman, *Gain-Cell Embedded DRAM under Cryogenic Operation-A First Study*, IEEE Trans. Very Large Scale Integr. Syst. **29**, 1319 (2021).

[43] M. Yamaoka, C. Yoshimura, M. Hayashi, T. Okuyama, H. Aoki, and H. Mizuno, *A 20k-Spin Ising Chip to Solve Combinatorial Optimization Problems with CMOS Annealing*, IEEE J. Solid-State Circuits **51**, 303 (2016).

[44] K. Yamamoto, K. Kawamura, K. Ando, N. Mertig, T. Takemoto, M. Yamaoka, H. Teramoto, A. Sakai, S. Takamaeda-Yamazaki, and M. Motomura, *STATICA: A 512-Spin 0.25M-Weight Annealing Processor with an All-Spin-Updates-at-Once Architecture for Combinatorial Optimization with Complete Spin-Spin Interactions*, IEEE J. Solid-State Circuits **56**, 165 (2021).

[45] D. Jiang, X. Wang, Z. Huang, Y. Yang, and E. Yao, *A Network-on-Chip-Based Annealing Processing Architecture for Large-Scale Fully Connected Ising Model*, IEEE Trans. Circuits Syst. I Regul. Pap. **70**, 2868 (2023).

[46] T. Takemoto, K. Yamamoto, C. Yoshimura, M. Hayashi, M. Tada, H. Saito, M. Mashimo, and M. Yamaoka, *4.6 A 144Kb Annealing System Composed of 9×16Kb Annealing Processor Chips with Scalable Chip-to-Chip Connections for Large-Scale Combinatorial Optimization Problems*, in *2021 IEEE International Solid- State Circuits Conference (ISSCC)* (IEEE, 2021), pp. 64–66.

[47] N. Dattani, S. Szalay, and N. Chancellor, *Pegasus: The Second Connectivity Graph for Large-Scale Quantum Annealing Hardware*, Preprint at http://arxiv.org/abs/1901.07636 (2019).

[48] K. Boothby, A. D. King, and J. Raymond, Zephyr Topology of D-Wave Quantum Processors,

2021.

[49] S. Choi, Y. Salamin, C. Roques-Carmes, R. Dangovski, D. Luo, Z. Chen, M. Horodynski, J. Sloan, S. Z. Uddin, and M. Soljacic, *Photonic Probabilistic Machine Learning Using Quantum Vacuum Noise*, Preprint at https://arxiv.org/abs/2403.04731v1 (2024).

[50] S. Luo, Y. He, B. Cai, X. Gong, and G. Liang, *Probabilistic-Bits Based on Ferroelectric Field-Effect Transistors for Probabilistic Computing*, IEEE Electron Device Lett. **44**, 1356 (2023).

[51] V. K. Joshi, P. Barla, S. Bhat, and B. K. Kaushik, *From MTJ Device to Hybrid CMOS/MTJ Circuits: A Review*, IEEE Access **8**, 194105 (2020).

[52] H. Lee, F. Ebrahimi, P. K. Amiri, and K. L. Wang, *Design of High-Throughput and Low-Power True Random Number Generator Utilizing Perpendicularly Magnetized Voltage-Controlled Magnetic Tunnel Junction*, AIP Adv. **7**, 055934 (2017).

[53] D. Vodenicarevic et al., *Low-Energy Truly Random Number Generation with Superparamagnetic Tunnel Junctions for Unconventional Computing*, Phys. Rev. Appl. **8**, 054045 (2017).

[54] K. Y. Camsari, M. M. Torunbalci, W. A. Borders, H. Ohno, and S. Fukami, *Double-Free-Layer Magnetic Tunnel Junctions for Probabilistic Bits*, Phys. Rev. Appl. **15**, 044049 (2021).

[55] K. Y. Camsari, S. Salahuddin, and S. Datta, *Implementing P-Bits with Embedded MTJ*, IEEE Electron Device Lett. **38**, 1767 (2017).

[56] A. Fukushima, T. Yamamoto, T. Nozaki, K. Yakushiji, H. Kubota, and S. Yuasa, *Recent Progress in Random Number Generator Using Voltage Pulse-Induced Switching of Nano-Magnet: A Perspective*, APL Mater. **9**, 030905 (2021).

[57] C. Safranski, J. Kaiser, P. Trouilloud, P. Hashemi, G. Hu, and J. Z. Sun, *Demonstration of Nanosecond Operation in Stochastic Magnetic Tunnel Junctions*, Nano Lett. **21**, 2040 (2021).

[58] K. Hayakawa, S. Kanai, T. Funatsu, J. Igarashi, B. Jinnai, W. A. Borders, H. Ohno, and S. Fukami, *Nanosecond Random Telegraph Noise in In-Plane Magnetic Tunnel Junctions*, Phys. Rev. Lett. **126**, 117202 (2021).

[59] L. Soumah, L. Desplat, N.-T. Phan, A. S. El Valli, A. Madhavan, F. Disdier, S. Auffret, R. Sousa, U. Ebels, and P. Talatchian, *Nanosecond Stochastic Operation in Perpendicular Superparamagnetic Tunnel Junctions*, Preprint at https://arxiv.org/abs/2402.03452 (2024).

[60] L. Schnitzspan, M. Kläui, and G. Jakob, *Nanosecond True-Random-Number Generation with Superparamagnetic Tunnel Junctions: Identification of Joule Heating and Spin-Transfer-Torque Effects*, Phys. Rev. Appl. **20**, 024002 (2023).

[61] Y. Shao, S. L. Sinaga, I. O. Sunmola, A. S. Borland, M. J. Carey, J. A. Katine, V. Lopez-Dominguez, and P. K. Amiri, *Implementation of Artificial Neural Networks Using Magnetoresistive Random-Access Memory-Based Stochastic Computing Units*, IEEE Magn. Lett. **12**, 4501005 (2021).

[62] K. Kobayashi, W. A. Borders, S. Kanai, K. Hayakawa, H. Ohno, and S. Fukami, *Sigmoidal Curves of Stochastic Magnetic Tunnel Junctions with Perpendicular Easy Axis*, Appl. Phys. Lett. **119**, 132406 (2021).

[63] I. Hen, J. Job, T. Albash, T. F. Rønnow, M. Troyer, and D. A. Lidar, *Probing for Quantum Speedup in Spin-Glass Problems with Planted Solutions*, Phys. Rev. A - At. Mol. Opt. Phys. **92**, 1 (2015).

[64] A. Biere, M. Heule, H. Van Maaren, and T. Walsh, *Handbook of Satisfiability: Volume 185 Frontiers in Artificial Intelligence and Applications* (IOS Press, 2009).

[65] Y. Shao, V. Lopez-Dominguez, N. Davila, Q. Sun, N. Kioussis, J. A. Katine, and P. Khalili Amiri, *Sub-Volt Switching of Nanoscale Voltage-Controlled Perpendicular Magnetic Tunnel Junctions*, Commun. Mater. **3**, 87 (2022).